\documentclass[prl,twocolumn,superscriptaddress,showpacs]{revtex4}

\usepackage{amsmath,amssymb,graphicx,color,bm}

\begin{document}

\title{Perceptrons with Hebbian learning based on wave ensembles in spatially patterned potentials}

\author{T. Espinosa-Ortega}
\affiliation{Division of Physics and Applied Physics, Nanyang Technological University 637371, Singapore}

\author{T. C. H. Liew}
\affiliation{Division of Physics and Applied Physics, Nanyang Technological University 637371, Singapore}

\date{\today}

\begin{abstract}
A general scheme to realize a perceptron for hardware neural networks is presented, where multiple interconnections are achieved by a superposition of Schr\"odinger waves. Spatially patterned potentials process information by coupling different points of reciprocal space. The necessary potential shape is obtained from the Hebbian learning rule, either through exact calculation or construction from a superposition of known optical inputs. This allows implementation in a wide range of compact optical systems, including: 1) any non-linear optical system; 2) optical systems patterned by optical lithography; and 3) exciton-polariton systems with phonon or nuclear spin interactions.
\end{abstract}

\maketitle

Neural networks exploit massive interconnectivity to become highly efficient at certain tasks, such as classification, and pattern recognition~\cite{Rojas1996,Graupe2013}. While biological neurons may operate individually on millisecond timescales, their simultaneous connection to several thousands of other neurons allows a parallelization of tasks far beyond the capabilities of CMOS logic. Naturally, this observation has motivated research into artificial neural networks, including hardware implementations~\cite{Misra2010}. Electrically connected systems have been based on  inorganic synapses~\cite{Ohno2011}, spiking silicon circuits~\cite{Indiveri2011,Poon2011}, feedback on Bose-Einstein condensates~\cite{Brynes2013}, memristors~\cite{Krzysteczko2012} or spin torque devices~\cite{Locatelli2014} such as nanomagnetic spin switches~\cite{Diep2014}. Optical systems can exploit vector matrix multiplication~\cite{Ohta1990} or fractional fourier transforms~\cite{Barshan2002}.

In models of neuron behaviour or in artificial neural networks, an individual neuron gives an output given by:
\begin{equation}
\xi^{out}_i=f\left(\sum_j w_{ij} \xi^{in}_j\right)
\label{eq:NN}
\end{equation}
where $f$ is some function depending on the sum of inputs from several input neurons, $\xi^{in}_j$, each multiplied with a different weight $w_{ij}$. The weights represent the ``knowledge'' in the system and are adjusted in a ``learning'' process or set externally for the desired function.

Given Eq.~\ref{eq:NN}, an artificial neural network requires three ingredients: a) A large number of interconnections between different neurons; b) The possibility of different weights for each interconnection; c) Flexibility in the choice of weights to allow learning. A two-layer network with one set of inputs and one layer of output neurons is already a powerful unit, known as a perceptron, capable of the linear classification of data and pattern recognition. However, a small device with $35$ inputs and $10$ outputs already requires $350$ independent weight connections. In electrical devices the engineering of so many connections is a serious task. Optical designs can benefit from the overlapping of several different light rays to build these connections; however, the controlled weighting of connections often requires separate electronic connections or bulky spatial light modulators.


Here a simple scheme of a perceptron is demonstrated, based on the superposition of an ensemble of waves. A model based on the Schr\"odinger equation illustrates the wide applicability of the scheme, which is compatible with a range of micron-sized solid-state implementations where a spatially varying potential is available. Examples include: 1) nonlinear optical systems; 2) systems microstructured by optical lithography; and 3) exciton-polariton systems with (i) acoustic phonon interactions or (ii) nuclear spin interactions. In general, the weights in the system can be calculated using the Hebbian learning rule or constructed via a superposition of optical waves. In examples 1 and 2, the weights are then written with a fixed potential. In example 3, the change (plasticity) of the effective potential, under a given set of inputs and outputs, allows a direct demonstration of Hebbian learning in which the system adapts automatically to perform the desired network function. The advantage of the proposed general scheme is that it is not necessary to construct or control each weight individually; it is only necessary to provide the appropriate stimulus as a combination of known input and output waves. Both permanent weightings for repetitive tasks as well as reconfigurable networks are possible. For illustration, the task of pattern recognition is considered. Operation on ultrafast (picosecond) timescales is expected.

\begin{figure}[th]
\includegraphics[width=.5\textwidth]{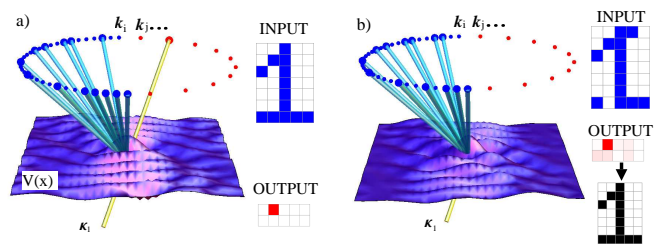}
\caption{a) \textit{Training phase}. A $35$ pixel input set corresponding to number one, is activated (blue rays) simultaneously with the desired output (red ray). The position of the activated output $\kappa_1$ identifies the number one. b) \textit{Operation phase}. The input set represents a digit with defects and there is no driven field over the output. Scattering with the potential nevertheless gives the maximum output corresponding to $|\psi_{\kappa_1}|^2$ and the number can be recognized.}
\label{Fig1}
\end{figure}

{\it General Scheme.---} Let us consider the 2D Schr\"odinger equation for particles with a wavefunction $\psi(\mathbf{x})$ moving in a spatially varying potential $V(\mathbf{x})$:
\begin{equation}
i\hbar\frac{\partial\psi}{\partial t}=\left(\hat{E}-i\hat{\Gamma}+V(\mathbf{x})\right)\psi+F(\mathbf{x})
\label{eq:Schrodinger}
\end{equation}
where $\hat{E}$ represents the kinetic energy and we allowed for driving ($F$) and decay terms ($\Gamma$), corresponding to a non-equilibrium system. The kinetic energy is measured from the energy associated with the driving field energy. Let us imagine a potential of the form:
\begin{equation}
V(\mathbf{x})=\sum_{ij}V_{ij}\cos\left(\left(\mathbf{k}_i-\mathbf{k}_j\right).\mathbf{x}\right)\label{eq:Vx}
\end{equation}
where the spatially dependent part of the wavefunction and driving terms can be decomposed into Fourier components in reciprocal space:
\begin{align}
\psi(\mathbf{x})=\frac{1}{\sqrt{S}}\sum_i\psi_ie^{i\mathbf{k}_i.\mathbf{x}}\label{eq:psix}\\
F(\mathbf{x})=\frac{1}{\sqrt{S}}\sum_iF_ie^{i\mathbf{k}_i.\mathbf{x}}
\label{Fx}
\end{align}
where $S$ is the area of the system and we restrict the driving field to wavevectors of the same magnitude, such that all states in the system have the same energy (which is conserved). Substituting Eqs.~\ref{eq:Vx}-\ref{Fx} into Eq.~\ref{eq:Schrodinger} and collecting terms oscillating at the same wavevector, we obtain:
\begin{equation}
i\hbar\frac{\partial \psi_i}{\partial t}=\left(E_0-i\Gamma_0\right)\psi_i+\sum_{j}V_{ij}\psi_j+F_i
\label{eq:Schrodinger2}
\end{equation}
where $E_0$ and $\Gamma_0$ are the energy and decay rates at $|\mathbf{k}_i|$ (now the same for all particles). We divide reciprocal space into a half containing driven wavevectors, which will characterize a vector of inputs in the system, $F_1$, $F_2$, etc, and a half for the output. The amplitudes $F_i\in(0,F_0)$ can take one of two values each, representing a binary input (see illustration in Fig.~\ref{Fig1}a).

Assuming that $V_{ij}$ is small, the steady state values in the input half of reciprocal space are:
\begin{equation}
\psi_i=-\frac{F_i}{E_0-i\Gamma}
\label{psi1}
\end{equation}
States in the output half of reciprocal space are reached via scattering with $V_{ij}$, giving the amplitudes:
\begin{equation}
\psi_i=-\frac{\sum_{ij}V_{ij}\psi_j}{E_0-i\Gamma}=\frac{\sum_{j}V_{ij}F_j}{\left(E_0-i\Gamma\right)^2}
\label{psiout}
\end{equation}
In other words the output intensities, $|\psi_i|^2$, are a function of $\sum_{j}V_{ij}F_j$, where $F_j$ are the inputs and $V_{ij}$ are the weight connections, as required for a neural network.

{\it Hebbian Learning.---} For a network to be useful, the weights $V_{ij}$ must be chosen to give the desired network function. Under Hebbian learning, neuron connection weights are increased when input and output neurons fire simultaneously. This allows training of the network where an input vector set is applied and the desired output state is simultaneously activated. Repeating the process over a ``training set'' of input vectors allows the system to learn how to distinguish different inputs, its ability being encoded in neuron connection weights of the form:
\begin{equation}
V_{ij}\propto\sum_{\{v\}}F_i^{\{v\}}F_j^{\{v\}}
\label{eq:HebbianLearning}
\end{equation}
where we sum over different vectors in the training set, which are labelled by the index $v$. Here the index $i$ represents vectors in the input half of reciprocal space. To activate states in the output half of reciprocal space, we apply the driving field represented by $F_j^{\{v\}}$.

Note that the driving field intensity for each training step is given by:
\begin{align}
&\left|F_i^{\{v\}}e^{i\mathbf{k}_i.\mathbf{x}}+F_j^{\{v\}}e^{i\mathbf{k}_j.\mathbf{x}}\right|^2\notag\\
&\hspace{10mm}=2F_i^{\{v\}}F_j^{\{v\}}\cos\left(\left(\mathbf{k}_i-\mathbf{k}_j\right).\mathbf{x}\right)+\mathrm{const.}
\label{eq:F}
\end{align}
Consequently, we seek a mechanism of varying the potential $V(\mathbf{x})$ in proportion to the driving field intensity during the training phase. This gives a field with the same form as in Eq.~\ref{eq:Vx} with weights $V_{ij}$ following the Hebbian learning rule (Eq.~\ref{eq:HebbianLearning}). The contribution of the constant in Eq.~\ref{eq:F} would also give a constant shift of the potential, however, this can be fully compensated by varying the energy of all the particles in the system (determined by the driving field frequency).

\begin{figure}[tb]
\includegraphics[width=.45\textwidth]{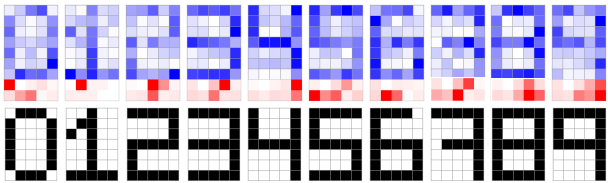}
\caption{Pattern recognition of numbers. Top-row: Input characters containing several errors. Middle-row: Light scattered by the potential is mapped to different outputs. Bottom-row: Character recognized by the system.}
\label{Fig3}
\end{figure}

As an illustration of Hebbian learning, we demonstrate the task of pattern recognition, where the potential is constructed from a superposition of known inputs and outputs from a training set of numerical characters ranging from $0-9$. Fig.~\ref{Fig3} shows that input characters (top row) are scattered to particular output wavevectors. In this way the system recognizes the characters (bottom row), even if the inputs contain multiple errors.

A key ingredient of the scheme is that the required potential has a spatial pattern proportional to the intensity of a known optical field. In the rest of the manuscript we consider different possible physical realizations of the potential. All cases follow the general recipe outlined above and allow direct reproduction of Fig.~\ref{Fig3}.

{\it 1) Nonlinear Optical Systems.}--- A wide variety of nonlinear optical systems are described by the nonlinear Schr\"odinger equation, identical to Eq.~\ref{eq:Schrodinger} with an additional term $\alpha|\psi|^2\psi$. If we imagine that the system can be excited by two different frequencies, $\omega_1$ and $\omega_2$, giving field components $\psi_1$ and $\psi_2$, then we can use the frequency $\omega_1$ for creating a potential of the form $\alpha|\psi_1|^2$ which is experienced by $\psi_2$. Each frequency component should again be decomposed in reciprocal space and the potential can be written either by direct calculation of the necessary optical field or by cycling over the input and output superpositions in the training set. The advantage of an optically induced potential~\cite{Amo2010} is that it can be changed at will, allowing rapid reconfiguration of the network function. Ideally, a system has multiple resonant modes to allow the efficient injection at different frequencies. Alternatively, one could make use of excitation of two orthogonal polarizations, provided there is some cross nonlinear interaction. The universality of the nonlinear Schr\"odinger equation implies that realization in spinor Bose-Einstein condensates and nanoparticle arrays~\cite{Noskov2012} could also be arranged.

{\it 2) Optical Lithography---} For repetitive applications, it may be desirable to make use of a permanent potential. These could be achieved by optical lithography, where the effective potential (refractive index) of a material is engineered in a thin film~\cite{OLithography} with a pattern dependent on its exposure to an incident optical field. The required pattern can again be constructed from a superposition of inputs $\{\textbf{k}_i\}$ and desired outputs $\kappa_1$ (see Fig.~\ref{Fig1}). In the simplest case the material structure resulting from optical lithography is varied between one of two values, representing a digitized potential $V(\textbf{r})\rightarrow(V_0,-V_0)$. The act of digitizing in real space, does not alter the relative weights of the relevant Fourier components of the potential such that the correct output field still obtained.

{\it 3) Semiconductor Microcavities.---} Exciton-Polaritons are particles that appear in semiconductor microcavities due to the strong coupling of quantum well excitons with cavity photons \cite{Microcavities}. Their injection into the system can be controlled via optical excitation and non-linear interactions between polaritons have been considered for building optical circuits based on binary logic architectures~\cite{Adrados2011,Ballarini2013,EspinosaOrtega2013}. Neural network architectures were not yet considered in this field, although polaritons are indeed good candidates for operating with variable potentials. For example, the injection of many polaritons introduces an effective potential due to polariton-polariton repulsion, which allows optical engineering of the potential landscape~\cite{Amo2010}, which is compatible with the scheme outlined in (1). Exciton-polariton systems offer further opportunities due to the existence of acoustic phonons and nuclear spin polarizations, which can provide a potential much longer lived than the typical picosecond scale of polariton dynamics.

{\it 3i) Acoustic Phonons.---} The interaction between polaritons and acoustic phonons is described by the Fr\"ohlich Hamiltonian~\cite{Takagahara85,Piermarocchi1996}:
\begin{equation}
\mathcal{H}_p=X\sum_{\mathbf{\mathfrak{q}},\mathbf{k}}\left(G_\mathbf{\mathfrak{q}}\hat{b}_\mathbf{\mathfrak{q}}\hat{a}^\dagger_{\mathbf{k}+\mathbf{\mathfrak{q}}}\hat{a}_\mathbf{k}+G^*_\mathbf{\mathfrak{q}}\hat{b}^\dagger_\mathbf{\mathfrak{q}}\hat{a}_{\mathbf{k}+\mathbf{\mathfrak{q}}}\hat{a}^\dagger_\mathbf{k}\right)
\label{Hp}
\end{equation}
where $\hat{a}_\mathbf{k}$ and $\hat{b}_\mathbf{\mathfrak{q}}$ are polariton and field operators in reciprocal space, respectively; $X$ is the excitonic fraction (Hopfield coefficient), which is the same for all polaritons given that they all have the same anergy; $\mathbf{\mathfrak{q}}=(\textbf{q},q_z)$. The explicit form of the exciton-phonon scattering amplitude $G_\mathbf{\mathfrak{q}}$ is written in Ref.~\cite{Piermarocchi1996}.

The phonon scattering rate can be calculated using the Fermi-golden rule,
\begin{align}
W_{ij,q_z}&=\frac{2\pi}{\hbar}|\langle\textbf{k}_j,n_{\textbf{q}_{ij},q_z}+1|\mathcal{H}_p|n_{\textbf{q}_{ij},q_z},\textbf{k}_i\rangle|^2\varrho_s\delta_{E_{\textbf{k}_j,\textbf{q}_{ij}}-E_{\textbf{k}_i}}\notag\\
&=\frac{2\pi}{\hbar}|X G^*_{ij,q_z}|^2 \varrho_s n_i^{pol}(n_{ij}^{ph}+1)(n_j^{pol}+1)
\end{align}
where $\varrho_s$ is the phonon density of states and $n^{pol}$, $n^{ph}$ stand for the polariton and phonon densities respectively. We integrate the expression $2\pi|X G^*_{ij,q_z}|^2 \varrho_s$ over $q_z$ (from $-2\pi/L_z$ to $2\pi/L_z$, with $L_z$ quantum well width) to obtain the in plane component of the scattering rate $w_{ij}\approx w(q\rightarrow0)\equiv w$. The dynamics of the phonon density is then given by:
\begin{equation}
\hbar\frac{\partial n^{ph}_{ij}}{\partial t}= w n_i^{pol}(n_{ij}^{ph}+1)(n_j^{pol}+1)-\Gamma_\chi n^{ph}_{ij}
\end{equation}
In analogy to Eq.(\ref{psi1}), the polariton densities for the driven signals follow $n^{pol}_i=|F_i|^2/(E_0^2+\Gamma^2)$. This allows straightforward calculation of the phonon densities obtained during the training phase [Fig.\ref{Fig1}a]. The optical driving field stimulates the excitation of phonons with specific wavevectors. In GaAs based systems the phonon lifetime has been measured in the range of $100ns$~\cite{Maznev2013} for low frequencies, which can be arranged by choosing small wavevectors, $|\mathbf{k}_i|=0.5\mu m^{-1}$, in reciprocal space. Consequently, the phonons leave a long-lasting mark through the polariton-phonon interactions, which will allow the system to ``memorize'' the digit.
\begin{figure}[h]
\includegraphics[width=.48\textwidth]{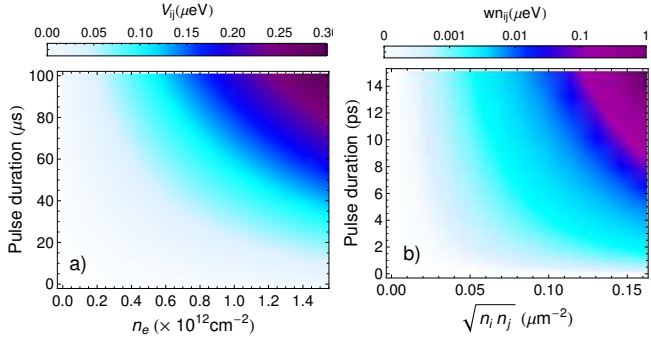}
\caption{a) Nuclear spin induced potential, given by Eq.\ref{eq:DNPVij}. b) Phonon-induced weight $w n_{ij}$ for a GaAs based system~\cite{Parameters}. The plots show the variation with the duration of application of input $k_i$ and output $k_j$, as well as the variation with electron density (a) and polariton density (b).}
\label{Fig2}
\end{figure}

During the operation phase, the output signals obey:
\begin{equation}
\hbar\frac{\partial n^{pol}_j}{\partial t}= \sum_i w n_i^{pol}(n_{ij}^{ph}+1)(n_j^{pol}+1)-\Gamma n^{pol}_j\label{eq:phononout}
\end{equation}
For a weak polariton density during operation and $\Gamma\gg\Gamma_\chi$, the polariton output density is given by:
\begin{equation}
n^{pol}_j=\frac{\sum_i w n_i^{pol}(n^{ph}_{ij}+1)}{\sum_iw n_i^{pol}(n^{ph}_{ij}+1)-\Gamma}\left( e^{(\sum_iw n_i^{pol}(n^{ph}_{ij}+1)-\Gamma)t}-1 \right)
\label{eq:phononSS}
\end{equation}
The analogy with Eq.~\ref{psiout} is most visible when $\Gamma$ is large. We verified that the obtained outputs allow the reproduction of Fig.~\ref{Fig3} for both GaAs and GaN based systems. Testing shorter phonon lifetimes revealed that $\Gamma_\chi$ is not a critical parameter for obtaining the correct outputs. $\Gamma_\chi$ is however important for determining the memory time of the system, which is on the one hand longer than the polariton lifetime to allow multiple repetitions and short enough to allow rapid reconfiguration of the device.


Within our theory we have neglected polariton-polariton scattering, which can in principle cause the redistribution of polaritons between modes on the circle in reciprocal space \cite{Leyder2007}. During training, this would have little effect as the dominant fields would still be those directly driven by the inputs, however, during operation it is implied that we operate in the low density regime. Considering an excitation polariton density of $1.5\times10^7 cm^{-2}$,  we obtain an output density of $6\times 10^6 cm^{-2}$ within a picosecond timescale. This output density is above the density that can be detected in experiments \cite{Stevenson2000,Langbein2002}, while also being below the density at which nonlinear effects become important \cite{Langbein2002,Brynes2014}.

{\it 3ii) Dynamic Nuclear Polarization.---} If during the training phase, the driving field is circularly polarized and the laser energy $E_p$ is increased so as to excite free electrons, then one can consider the dynamic polarization of nuclei. The hyperfine interaction between a single electron and single nuclear spin is given by the Hamiltonian~\cite{Gammon2001,Imamoglu2003,Erlingson2001}:
\begin{equation}
\mathcal{H}_\mathrm{hf}=\nu_0A|\Phi(\mathbf{R})|^2\left(\hat{I}_x\hat{S}_x+\hat{I}_y\hat{S}_y+\hat{I}_z\hat{S}_z\right)
\end{equation}
where $\nu_0$ is the unit cell volume and $A$ is the hyperfine coupling constant. $\Phi(\mathbf{R})$ represents the electron envelope function, evaluated at the position of the nuclear spin. $\hat{\mathbf{I}}$ and $\hat{\mathbf{S}}$ represent the nuclear and electron spin operators, respectively. An electron with spin polarized in say the $z$-direction can undergo a spin flip, transferring its spin to a nucleus. The average nuclear spin polarization in the system is determined by the rate equation:
\begin{equation}
\frac{d\langle I_z(\mathbf{x})\rangle}{dt}=\Gamma_\mathrm{hf}(\mathbf{x})\left(Q\langle S_z\rangle-\langle I_z(x)\rangle\right)-\frac{\langle I_z(\mathbf{x})\rangle}{\tau}
\label{dIz}
\end{equation}
where $\tau$ is the nuclear spin relaxation time, which takes values at least on the order of microseconds~\cite{Braun2006,Maletinsky2007} and even up to minutes~\cite{Krapf1990}. The quantity $Q=\frac{I(I+1)}{S(S+1)}$, where $I$ is the total nuclear spin and $S_z=1/2$ is the total spin of an electron. The hyperfine scattering rate $\Gamma_\mathrm{hf}$, was calculated for a semiconductor microcavity in Ref.~\cite{Liew2011} and is proportional to the electron density. Compared to the wavelength of polaritons, electrons do not move significant distances during their lifetime such that we can consider the hyperfine scattering rate as being proportional to the optical field intensity. Solving Eq.(\ref{dIz}):
\begin{equation}
\langle I_z(\mathbf{x})\rangle=\frac{\Gamma_\mathrm{hf} Q S_z (1-e^{-(\Gamma_\mathrm{hf}+\tau^{-1})t})}{\Gamma_\mathrm{hf}+\tau^{-1}}
\end{equation}
For small quantum wells ($L_z\sim10nm$) and electron density around we have that $\Gamma_\mathrm{hf}\ll\tau^{-1}$, therefore we can approximate the above expression as $\langle I_z(\mathbf{x})\rangle=\Gamma_\mathrm{hf} Q S_z \tau(1-e^{-t/\tau})$. Consequently, the polariton potential is given by $V(\mathbf{x})=X A\langle I_z(\mathbf{x})\rangle\propto |F_0|^2$.

The potential that the nuclear spin polarization induces on the polaritons is given by
\begin{equation}
V_{ij}=\frac{8\tau X A^3 Q S_z  m_e \nu_0^2 I(I+1)}{3 \hbar^3 V L_z E_p^2}F_i F_j\label{eq:DNPVij}
\end{equation}
with $E_p=E_c(|\textbf{k}|)$, the energy of the cavity mode at the pump wave vector. The nuclear spin induced potential is shown in Fig.~\ref{Fig2}b, for different exposure times and electron densities.

For the considered densities, the nuclear spin induced potentials correspond to nuclear spin polarizations of around $1\%$.  While nuclear spin polarizations up to $3.3\%$ and $10\%$ were observed in quantum wells \cite{Barrett1994} and quantum dots \cite{Lai2006}, respectively, and higher electron densities were considered in the literature~\cite{Laussy2010}, the small nuclear spin induced potential is in fact sufficient for our task. If one compares to the strength of disorder, which is around $10\mu$eV in state-of-the-art samples, then the nuclear spin induced potential may seem insignificant. However, for the determination of the scattered polariton intensity one must examine the Fourier components of the potentials. In reciprocal space, the nuclear spin induced potentials are localized at specific points while a random disorder potential is spread widely. In fact, if one compares the Fourier amplitude $V_{ij}$ to the typical Fourier amplitude $V_\mathrm{disorder}$ of a Gaussian correlated disorder potential, one finds: $V_{ij}/V_\mathrm{disorder}=V_{ij}\sqrt{S}e^{(k_i-k_j)^2\sigma^2/r}/(\sigma V_\mathrm{rms})$, where $S$ is the system area, $\sigma$ the disorder correlation length and $V_\mathrm{rms}$ is the root mean squared amplitude of disorder. After all training steps are completed $V_{ij}\sim2\mu$eV. Typically, microcavities have a larger disorder than in Ref.~\cite{Zajac2012}, characterized by a height of $0.1$meV or root mean squared variation $V_\mathrm{rms}\approx10\mu$eV and correlation length $\sigma\approx1\mu$m. Taking these values, $k_i-k_j=1\mu$m$^{-1}$, and a wide excitation area $S=200\mu$m$^2$, one finds that the ratio of scattered polariton intensity in reciprocal space by the nuclear spin induced potential and disorder potential is $n_j/n_\mathrm{disorder}=\left(V_{ij}/V_\mathrm{disorder}\right)^2\sim1600$. Consequently the nuclear spin induced potential has a much stronger effect on polariton scattering than disorder, and would be clearly visible experimentally on picosecond time scales \cite{Langbein2002}.

Substituting Eq.~\ref{eq:DNPVij} into Eq.~\ref{psiout}, we verified that Fig.~\ref{Fig3} was reproducible for typical parameters~\cite{Parameters}.




{\it Conclusion.---} We developed a general scheme to construct perceptrons using wave ensembles, appropriate to a wide range of optical systems. Unlike other hardware implementations of neural networks, our general scheme avoids the need to store and control the weight of each network connection, $w_{ij}$, in an independent feedback loop. A large number of weights are encoded in the form of the spatially structured potential, which is constructed from exposure to a known set of training conditions.

We considered three classes of example systems: 1) nonlinear optical systems, 2) systems patterned by optical lithography, and 3) exciton-polariton systems. In nonlinear optical systems, the optically induced potential can be rapidly reconfigured, while optical lithography allows the engineering of a permanent network for a repetitive task. The exciton-polariton systems allow network weights to be automatically chosen by phonons or nuclear spins inside the system, allowing the system to learn in response to a given set of stimuli. Due to the wide availability of suitable systems, we believe this approach will lead to ultra-fast and compact hardware implementations of neural networks as well as a new platform for machine learning.

We thank I.A. Shelykh for discussion and acknowledge support from the NTU grant $M4080853.110.706022$.


\begin{thebibliography}{99}

\bibitem{Rojas1996}
R. Rojas, {\it Neural Networks}, Springer-Verlag, New York, (1996).

\bibitem{Graupe2013}
D. Graupe, {\it Principles of Artificial Neural Networks}, World Scientific (2013).

\bibitem{Misra2010}
Misra \& I. Saha, Neurocomputing, {\bf 74}, 239 (2010).

\bibitem{Ohno2011}
T. Ohno, T. Hasegawa, T. Tsuruoka, K. Terabe, J. K. Gimzewski and M. Aono, Nature Mater. \textbf{10}, 591 (2011).

\bibitem{Indiveri2011}
G Indiveri, {\it et al.}, Front Neurosci., {\bf 5}, 73 (2011).

\bibitem{Poon2011}
C S Poon and K Zhou, Front Neurosci., {\bf 5}, 108 (2011).

\bibitem{Brynes2013}
T Brynes, S Koyama, K Yan, \& Y Yamamoto, Sci. Rep., {\bf 3}, 2531 (2013).

\bibitem{Krzysteczko2012}
P. Krzysteczko, J. Munchenberger, M. Schafers, G. Reiss, \& A. Thomas, Adv. Mater., {\bf 24}, 762 (2012).

\bibitem{Locatelli2014}
N. Locatelli, V. Cros, \& J. Grollier, Nature Mater., {\bf 13}, 11 (2014).

\bibitem{Diep2014}
V. Q. Diep, B. Sutton, B. Behin-Aein, \& S. Datta, arXiv:1404.2654 (2014).

\bibitem{Ohta1990}
J. Ohta, K. Kojima, Y. Nitta, S. Tai, \& K. Kyuma, Opt. Lett., {\bf 15}, 1362 (1990).

\bibitem{Barshan2002}
B. Barshan and B. Ayrulu, Neural Net. \textbf{15}, 131 (2002).

\bibitem{Amo2010}
A. Amo, {\it et al.}, Phys. Rev. B, {\bf 82}, 081301(R) (2010).

\bibitem{Noskov2012}
R E Noskov, P A Below, \& Y S Kivshar, Phys. Rev. Lett., {\bf 108}, 093901 (2012).



\bibitem{Adrados2011}
C. Adrados, {\it et al.}, Phys. Rev. Lett., {\bf 107}, 146402 (2011).

\bibitem{Ballarini2013}
D. Ballarini, {\it et al.}, \&, Nature Comm., {\bf 4}, 1778 (2013).

\bibitem{EspinosaOrtega2013}
T. Espinosa-Ortega \& T. C. H. Liew, Phys. Rev. B, {\bf 87}, 195305 (2013).


\bibitem{OLithography}
A. Faraon, et al., Appl. Phys. Lett., {\bf 92}, 043123 (2008).

\bibitem{Microcavities}
A. V. Kavokin, J. J. Baumberg, G. Malpuech, and F. P. Laussy,
Microcavities (Oxford University Press, New York, 2007).

\bibitem{Takagahara85}
T. Takagahara, Phys. Rev. B, {\bf 31}, 6552 (1985).

\bibitem{Piermarocchi1996}
C. Piermarocchi, F. Tassone, V. Savona, A. Quattropani, P. Schwendimann, Phys. Rev. B, {\bf 53}, 15834 (1996).

\bibitem{Maznev2013}
A. A. Maznev, {\it et al.}, Appl. Phys. Lett., {\bf 102}, 041901 (2013).

\bibitem{Leyder2007}
C. Leyder, {\it et al.}, Phys. Rev. Lett., \textbf{99}, 196402 (2007).

\bibitem{Stevenson2000}
R. M. Stevenson, {\it et al.}, Phys. Rev. Lett., \textbf{85}, 3680 (2000).

\bibitem{Langbein2002}
W. Langbein \& J. M. Hvam, Phys. Rev. Lett., \textbf{88}, 047401 (2002).

\bibitem{Brynes2014}
T. Brynes, N. Y. Kim, \& Y. Yamamoto, Nature Phys., \textbf{10}, 803 (2014).

\bibitem{Gammon2001}
D. Gammon, {\it et al.}, Phys. Rev. Lett., {\bf 86}, 5176 (2001).

\bibitem{Imamoglu2003}
A. Imamoglu, E. Knill, L. Tian, \& P. Zoller, Phys. Rev. Lett., {\bf 91}, 017402 (2003).

\bibitem{Erlingson2001}
S. I. Erlingsson, Y. V. Nazarov, \& V. I. Fal'ko, Phys. Rev. B, {\bf 64}, 195306 (2001).

\bibitem{Braun2006}
P. F. Braun, {\it et al.}, Phys. Rev. B, {\bf 74}, 245306 (2006).

\bibitem{Barrett1994}
S. E. Barrett, R. Tycko, L. N. Pfeiffer and K. W. West, Phys. Rev. Lett., \textbf{72}, 1368 (1994).

\bibitem{Maletinsky2007}
P. Maletinsky, A. Badolato, \& A. Imamoglu, Phys. Rev. Lett., {\bf 99}, 056804 (2007).

\bibitem{Krapf1990}
M. Krapf, {\it et al.}, Solid State Commun., {\bf 74}, 1141 (1990).

\bibitem{Liew2011}
T. C. H. Liew \& V. Savona, Phys. Rev. Lett., {\bf 106}, 146404 (2011).

\bibitem{Lai2006}
C W Lai, P Maletinsky, A Badolato, \& A Imamoglu, Phys. Rev. Lett., {\bf 96}, 167403 (2006).

\bibitem{Laussy2010}
F P Laussy, A V Kavokin, \& I A Shelykh, Phys. Rev. Lett., {\bf 104}, 106402 (2010).


\bibitem{Zajac2012}
J M Zajac, E Clarke, \& W Langbein, Appl. Phys. Lett., {\bf 101}, 041114 (2012).



\bibitem{Parameters}
$m_e=0.069m_0$, $m_h=0.18m_0$, $\Gamma=0.16$meV and $\Gamma_x=3.3\mu$eV, $E_0=0$, $\rho=5.3g/cm^3$, $u=5.37\times10^3m/s$, $a_B=5.8nm$, $a_e=-7eV$, $a_h=2.7eV$ $\nu_0=(5.65 {\AA})^3/2$, $A=90 \mu eV$, $Q=5$, $S_z=1$, $\tau=10ms$, $w=15neV$.


\end{thebibliography}
\end{document}